\documentclass[prb,twocolumn,msmath,superscriptaddress,amssymb,showpacs]{revtex4}

\usepackage{color} 
\usepackage{graphicx}
\usepackage{epstopdf}
\usepackage{bm}
\usepackage{mathtools}
\usepackage{amsmath}

\begin{document}

\DeclareGraphicsExtensions{.eps, .png, .jpg}
\bibliographystyle{prsty}

\title {Emergent Dirac carriers across a pressure-induced Lifshitz transition in black phosphorus}

\author{P. Di Pietro}
\affiliation{Elettra-Sincrotrone Trieste, Area Science Park, I-34012 Trieste, Italy} 

\author {M. Mitrano}
\affiliation{Department of Physics and Materials Research Laboratory, University of Illinois, Urbana, IL 61801, USA}

\author{S. Caramazza}
\affiliation{Dipartimento di Fisica, Universit\`a di Roma Sapienza,  P.le Aldo Moro 2, I-00185 Roma, Italy} 

\author{F. Capitani}
\affiliation{Dipartimento di Fisica, Universit\`a di Roma Sapienza,  P.le Aldo Moro 2, I-00185 Roma, Italy} 
\affiliation{Synchrotron SOLEIL, LÕOrme des Merisiers, Saint-Aubin, 91192 Gif-sur-Yvette, France}

\author{S. Lupi}
\affiliation{CNR-IOM and Dipartimento di Fisica, Universit\`a di Roma Sapienza,  P.le Aldo Moro 2, I-00185 Roma, Italy} 

\author{P. Postorino}
\affiliation{CNR-IOM and Dipartimento di Fisica, Universit\`a di Roma Sapienza,  P.le Aldo Moro 2, I-00185 Roma, Italy} 

\author{F. Ripanti}
\affiliation{Dipartimento di Fisica, Universit\`a di Roma Sapienza,  P.le Aldo Moro 2, I-00185 Roma, Italy} 

\author {B. Joseph}
\affiliation{Elettra-Sincrotrone Trieste, Area Science Park, I-34012 Trieste, Italy} 

\author {N. Ehlen}
\affiliation{Physikalishes Institut, Universit\"at zu K\"oln, Z\"ulpicher Strasse 77, 50937 K\"oln, Germany}

\author {A. Gr\"uneis}
\affiliation{Physikalishes Institut, Universit\"at zu K\"oln, Z\"ulpicher Strasse 77, 50937 K\"oln, Germany}

\author {A. Sanna}
\affiliation{Max Planck Institut f\"ur Microstrukturphysik, Weinberg 2, D-06120 Halle, Germany}

\author{G. Profeta}
\affiliation{CNR-SPIN and Dipartimento di Fisica, Universit\`a degli Studi di L'Aquila, Via Vetoio 10,  I-67100 L'Aquila, Italy}

\author{P. Dore}
\affiliation{CNR-SPIN and Dipartimento di Fisica, Universit\`a di Roma Sapienza,  P.le Aldo Moro 2, I-00185 Roma, Italy} 

\author{A. Perucchi}
\affiliation{Elettra-Sincrotrone Trieste, Area Science Park, I-34012 Trieste, Italy}

\date{\today}

\begin{abstract}

The phase diagrams of correlated systems like cuprates or pnictides high-temperature superconductors are characterized by a topological change of the Fermi surface under continuous variation of an external parameter, the so-called Lifshitz transition. However, the large number of low-temperature instabilities and the interplay of multiple energy scales complicate the study of this phenomenon. Here we first identify the optical signatures of a pressure-induced Lifshitz transition in a clean elemental system, black phosphorus. By applying external pressures above 1.5 GPa, we observe a change in the pressure dependence of the Drude plasma frequency due to the appearance of massless Dirac fermions. At higher pressures, optical signatures of two structural phase transitions are also identified. Our findings suggest that a key fingerprint of the Lifshitz transition in solid state systems, and in absence of structural phase transitions, is a discontinuity of the Drude plasma frequency due to the change of Fermi surface topology.

\end{abstract}
\maketitle

The Lifshitz transition, the change of the Fermi surface topology under variation of an external parameter \cite{lifshitz1960}, is a fundamental phenomenon in strongly correlated systems like the YBa$_2$Cu$_3$O$_{6+y}$\cite{norman10}, Bi$_2$Sr$_2$CaCu$_2$O$_{8+\delta}$ \cite{benhabib15} and Ba(Fe$_{1-x}$Co$_x$)$_2$As$_2$ \cite{liu10} superconductors, and is suspected to play a relevant role in determining their electronic properties. In these materials, it may induce a band flattening and increase the density of states close to the Fermi level, thus promoting high-temperature superconductivity\cite{volovik17}. However, these compounds also exhibit a variety of low-temperature phase transitions that can mask the thermodynamic and transport properties of a pure Lifshitz transition. As a result, the dynamical charge and current fluctuations across a Lifshitz transition are still poorly understood.

Several different Lifshitz transitions have been observed (or theoretically predicted) in elemental black phosphorus, as a function of doping\cite{kim15,ehlen18}, electric field \cite{liu15}, and pressure \cite{xiang15,li17}. BP is an attractive material for electronic applications due to its very high electron mobility (10$^4$ cm$^2$V$^{-1}$s$^{-1}$) and the presence of a direct, tunable infrared band gap \cite{qiao14}.
Its A17 orthorhombic structure is extremely anisotropic, with grooves oriented along the so-called zig-zag direction \cite{low14} (see Fig.1a). The band structure is parabolic along both the interlayer and the zig-zag directions, while along the armchair direction the dispersion is almost linear, thus allowing possible Dirac cones \cite{fei15,ehlen16}. 
Upon pressurization, both the structure and the electronic properties undergo dramatic changes and a Lifshitz transition occurs at a pressure $P_L=1.5$ GPa (see Fig.1b). At low pressures, the electronic band gap gradually closes until the valence and conduction bands touch at the Z point, before intersecting each other without hybridizing. Around $P_L$, four-fold degenerate Dirac points are formed, which are then evolving in both electron and hole-like Fermi pockets when the pressure is further increased\cite{gong17}. At $P>P_L$, the orthorombic (A17) structure becomes rhombohedral (A7) around 5 GPa and then simple cubic (sc) around 10 GPa \cite{jamieson63, Kikeg83, Akai89,scelta17}, where superconductivity also occurs \cite{kawamura85,zhang17}.

The occurrence of a pressure-induced Lifshitz transition in an elemental semiconductor provides a unique setting for the study of its electrodynamics in the absence of other low-temperature instabilities. Here, we address this topic by driving BP across the pressure-induced Lifshitz transition and studying its optical response with synchrotron-based infrared spectroscopy and first principles density functional theory (DFT) calculations. We identify for the first time the optical signatures of a pressure-induced topological Lifshitz transition in an elemental semiconductor. At a transition pressure $P_L=1.5$ GPa, BP evolves from a semiconductor to a Dirac semimetal by building up a plasma of massless charge carriers. At higher pressures, we observe the optical fingerprints of the two structural phase transitions occurring in the semimetal phase.

% Figure 1
\begin{figure}[t]
\begin{center}
\leavevmode
\includegraphics [width=\columnwidth]{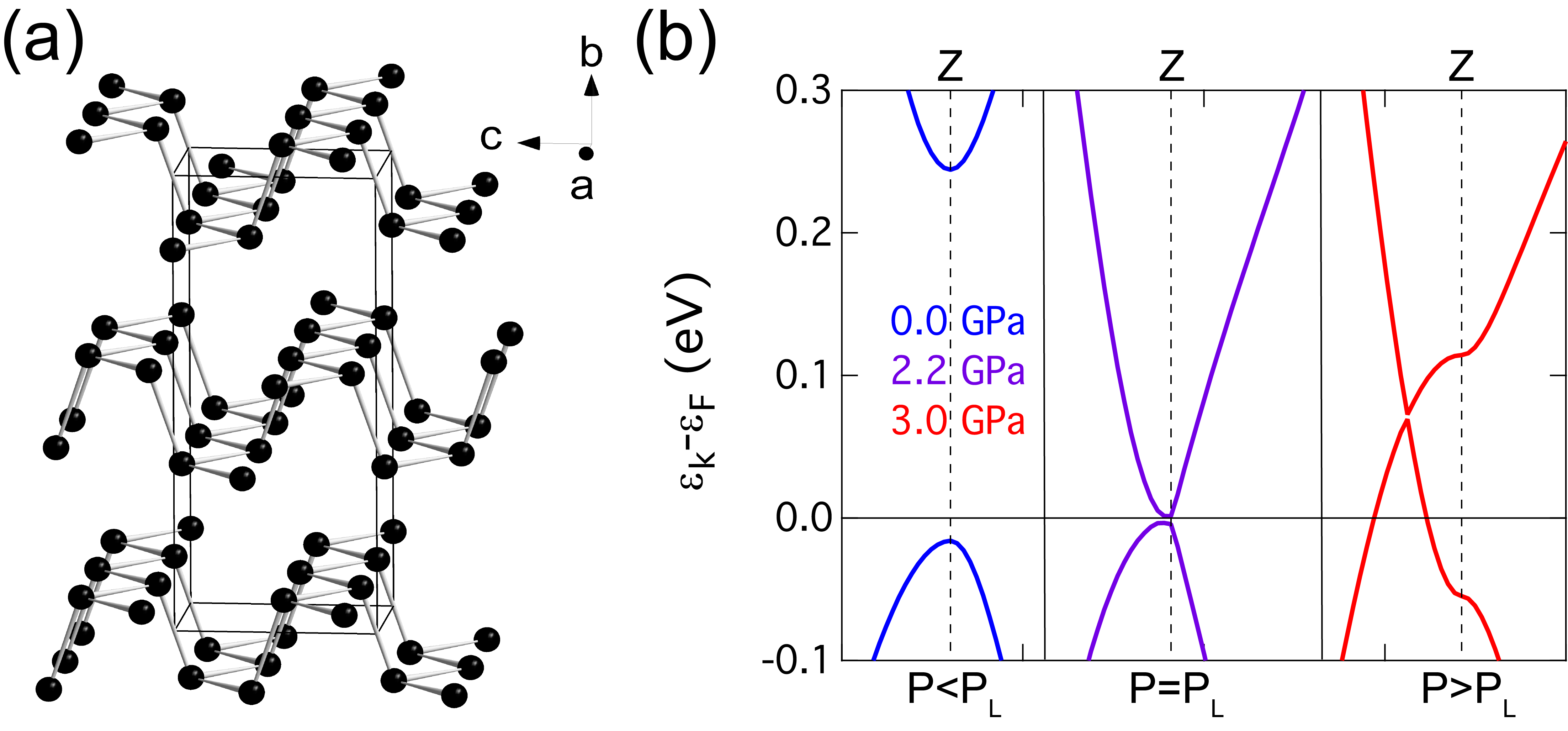}  
\end{center}
\caption{ (Color online) (a) Crystallographic unit cell of BP. (b) Valence and conduction bands of BP at the Z point (marked by a dashed line) across the pressure-induced Lifshitz transition. $P_L$ is the Lifshitz transition pressure, while the Z points are offset in momentum for clarity.}
\label{Fig1}
\end{figure}

Our infrared measurements under pressure have been performed at the SISSI infrared beamline\cite{lupi07} of the Elettra storage ring, with a Bruker 70v interferometer mated to a broadband infrared microscope. BP samples were obtained from two different providers (Smart-Elements and HQ Graphene) and cut for use in a Diamond Anvil Cell (DAC) with CsI as the pressure transmitting medium, yielding identical experimental results. The samples, oriented along the basal $ac$ plane (see Fig1a), were kept in contact with the diamonds in order to ensure a flat interface. Pressure was gauged through the ruby fluorescence technique\cite{mao86}.
The samples were mounted in two different DACs equipped with 1 mm and 0.4 mm culet diamonds respectively. The former allowed reliable measurements down to 100 cm$^{-1}$ for pressures up to 2.2 GPa, while the latter allowed up to 10.4 GPa. Light was polarized along the most conductive direction,  the $c$ (armchair) axis crystallographic direction. From the reflectivity data at the sample-diamond interface we retrieved the optical conductivity through Kramers-Kronig transformations \cite{Dressel,perucchi09}. All the measurements reported in this study were performed at room-temperature (RT).

%Figure 2
\begin{figure}[t]
\begin{center}
\leavevmode
\includegraphics [width=\columnwidth]{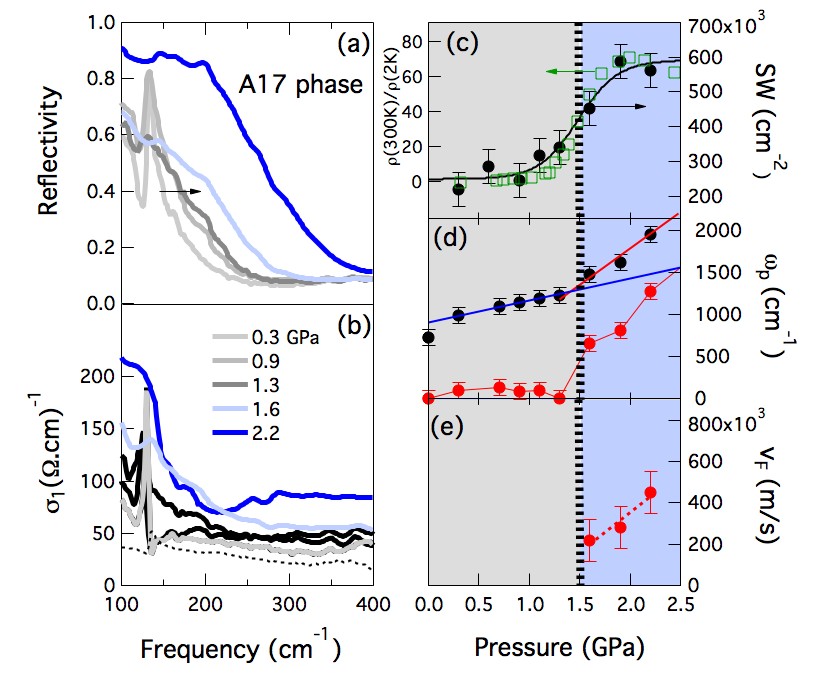}  
\end{center}
\caption{(Color online) Optical reflectivity at the sample-diamond interface (a) and real part of the optical conductivity (b) of orthorhombic BP measured along the $c$-axis polarization direction.  The ambient pressure optical conductivity is also reported for comparison as a black dashed line. (c) Spectral weight (SW) of $\sigma_1(\omega)$ integrated between 100 and 200 cm$^{-1}$. SW follows a phenomenological sigmoid behavior centered at $P_L=1.5$ GPa (dashed vertical line). The resistivity ratio $\rho(300K)/\rho(2K)$ from Ref. \onlinecite{xiang15} is also reported for comparison. (d) Plasma frequency $\omega_p$ (from Lorentz-Drude fitting) vs pressure. Black circles represent the bare $\omega_p$, while red circles mark the Dirac $\omega_p$ (see text) (e) Fermi velocity of Dirac carriers from equation (\ref{dirac}). The red dashed line is a guide to the eye.}
\label{Fig2}
\end{figure}

We report in Fig. \ref{Fig2} the infrared signatures of the pressure-induced Lifshitz transition in the orthorhombic phase of BP, the key experimental observation of this work. Above 200 cm$^{-1}$, the low pressure ($<1$ GPa) reflectivity $R(\omega)$  is approximately 0.1, and remains flat in the whole measured range, up to 8000 cm$^{-1}$. At 130 cm$^{-1}$, we detect a sharp peak assigned to the $B_{1u}$ phonon mode\cite{sugai85}. The real part of the optical conductivity $\sigma_1(\omega)$ is very low and consistent with semiconducting behavior at ambient pressure. As pressure is increased, both $R(\omega)$ and  $\sigma_1(\omega)$ are gradually enhanced. However, between 1.3 and 1.6 GPa, we observe an abrupt blue-shift of the reflectivity plasma edge that can be ascribed to the Lifshitz transition observed in ARPES \cite{kim15}. As pressure is further increased and the phonon becomes screened, a Drude-like absorption term appears due to the delocalization of charge carriers. More details about the pressure dependence of the $B_{1u}$ phonon are reported in the supplementary information.

A reliable, model-independent figure of merit for the pressure-induced metallization is the low frequency spectral weight\cite{Dressel}, $SW=\frac{120}{\pi}\int_{\Omega_1}^{\Omega_2}\sigma(\omega)d\omega$, integrated between $\Omega_1=100$ cm$^{-1}$ and $\Omega_2=200$ cm$^{-1}$. $\Omega_1$ is the low frequency limit of our data, while $\Omega_2$ is chosen in order to fully include the low frequency Drude term. As visible in Fig. \ref{Fig2}c, the pressure dependent SW follows a sigmoid trend centered at $P_L=1.5$ GPa. This behavior maps exactly onto the resistivity measurement from Ref. \onlinecite{xiang15} (see Fig. 2c), and was previously associated with the Lifshitz transition. 

\begin{figure*}[t]
\begin{center}
\leavevmode
\includegraphics [width=16cm]{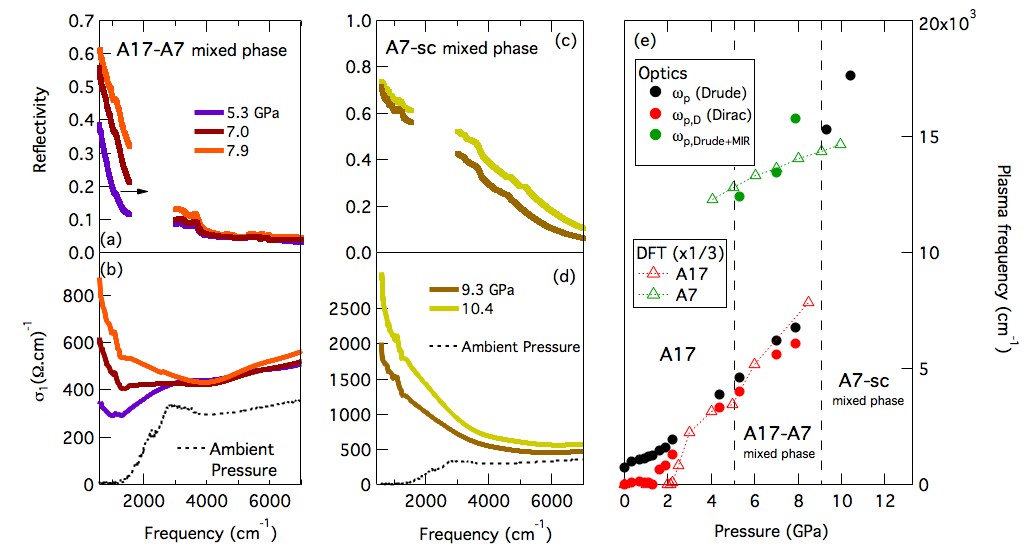}  
\end{center}
\caption{(Color online) Optical reflectivity at the sample-diamond interface (a) and real optical conductivity (b) of BP measured along the $c$-axis polarization direction in the A17-A7 mixed phase. (c)-(d) Same quantities in the A7-sc mixed phase. (e) Experimental (full circles) and theoretical (empty triangles) Drude plasma frequencies across the structural phase transitions. Black circles represent the bare Drude-Lorentz plasma frequency. The Dirac plasma frequency (red circles) is obtained by subtracting the plasma frequency of the thermally activated massive carriers. The sum in quadrature of the Drude and MIR band plasma frequencies is shown as green circles. Red and green triangles are the rescaled DFT plasma frequencies in the A17 and A7 phases, respectively.}
\label{Fig3}
\end{figure*}

In order to quantify the carrier density changes across the transition, we performed a Drude-Lorentz fitting of the data. The plasma frequency ($\omega_p$) associated with the free-carrier Drude term is reported in Fig. \ref{Fig2}d as a function of pressure. It is worth noting that a small, but sizeable RT conductivity in the order of few 10's ($\Omega\cdot\textrm{cm}$)$^{-1}$ is observed also at ambient pressure, i.e. in the semiconducting phase, and can be ascribed to the presence of thermally-activated carriers. Notably, the plasma frequency increases linearly with pressure, up to $P_L$. 

At the Lifshitz transition pressure $P_L$, $\omega_p$ increases more steeply at a rate of 900 cm$^{-1}$/GPa, almost three times the 260 cm$^{-1}$/GPa slope observed below $P_L$. The discontinuity in the plasma frequency slopes can be ascribed to the simultaneous presence of different fermions contributing to the conduction. Below $P_L$ only massive, thermally activated carriers contribute to the conduction. As the Dirac cone is formed above $P_L$, also Dirac-like fermions contribute to the Drude conductivity, thus leading to a combination of massive (Schr\"odinger-like)
and massless (Dirac) carriers. We evaluate the massive carrier contribution ($\omega_{p,S}$) at all pressures by extrapolating above $P_L$ the linear behavior of $\omega_p$ from below $P_L$. 
As a consequence, the massless Dirac contribution to the plasma frequency $\omega_{p,D}$ can be calculated at all pressures from $\omega_p^2=\omega_{p,S}^2+\omega_{p,D}^2$. The resulting pressure dependent $\omega_{p,D}$ is reported in Fig. \ref{Fig2}d.

The massless Dirac plasma frequency $\omega_{p,D}$ can be microscopically calculated as \cite{dassarma09}
\begin{equation}
\omega_{p,D}=\sqrt{\frac{e^2}{\hbar v_F}}\bigg(\frac{32\pi g_s g_v}{3}\bigg)^{1/6}n^{1/3}v_F, \label{dirac}
\end{equation}
where $g_s$ and $g_v$ are the spin and valley degeneracies ($g_s$=2 and $g_v$=2 in BP \cite{jiang15}). By making use of the experimental density of carriers determined from Hall effect measurements\cite{akiba17}, we can use Eq. \ref{dirac} to estimate the pressure-dependent Fermi velocity $v_F$ (reported in Fig. \ref{Fig3}c), and we find it to be around 2$\div4\cdot10^6$ m/s in good agreement with theoretical calculations \cite{gong17}.

By further increasing pressure well above the Lifshitz transition, BP undergoes two distinct structural phase transitions, from orthorombic (A17) to rhombohedral (A7), and from rhombohedral to simple cubic (sc), at about 5 and 10 GPa respectively \cite{jamieson63,Kikeg83,kawamura85}. Recent experiments hint to the presence of large regions of phase coexistence between the various structural phases \cite {guo17,joseph17,livas17,gupta17}. According to x-ray diffraction data from Ref. \onlinecite{joseph17}, performed on the same batch of samples used in this work, the A7 phase starts to appear above 5 GPa and coexists with A17 up to 10 GPa. Above this pressure, the A7 phase disappears, while the sc phase gradually sets in.

We report in Fig. \ref{Fig3}a-b, the optical properties across the A17-A7 phase transition to investigate how the mixed Drude responds to a structural phase transition. Under increasing pressure, the infrared reflectivity is enhanced and its plasma edge monotonically blue-shifts (see Fig. \ref{Fig3}e), while
the optical conductivity increases. The optical gap, located roughly at $\sim2000$ cm$^{-1}$ (see the ambient pressure $\sigma_1(\omega)$ reported in Fig. 3b for reference) is filled up. In this pressure range, the optical conductivity can be described by the combination of a Drude term and a mid-infrared (MIR) band (see supplemental material). The Drude plasma frequency grows linearly with pressure (black circles in Fig. \ref{Fig3}e) with the same 900 cm$^{-1}$/GPa slope observed at 1.5 GPa. With increasing pressure, the MIR band progressively coalesces into the Drude and becomes a second zero-frequency oscillator with a scattering rate $\gamma\sim$ 6000 cm$^{-1}$, i.e. much larger than the one associated to the massless carriers ($\gamma\sim$50$\div$500 cm$^{-1}$). Remarkably, when summed in quadrature ($\omega_{p,{\textrm{Drude+MIR}}}$), the two Drude terms exhibit a linear increase with the same slope of 900 cm$^{-1}$/GPa discussed above (see Fig. \ref{Fig3}e). 
Let us note here that in this pressure range the decomposition of $\omega_p$ in terms of $\omega_{p,D}$ and $\omega_{p,S}$ becomes relatively unimportant because of the dominance of the $\omega_{p,D}$ contribution (see Fig. 3e).

When entering the high pressure A7-sc mixed phase, the optical properties drastically change (Fig. \ref{Fig3}c-d). The reflectivity edge shifts to 8000 cm$^{-1}$, resulting in a greatly enhanced $\sigma_1(\omega)$. The two-bands electronic structure clearly identified in the A17-A7 mixed phase appears now to be merged into one single Drude term with $\omega_{p}\sim15000$ cm$^{-1}$, roughly corresponding to the $\omega_{p,{\textrm{Drude+MIR}}}$ term introduced before to describe the A17-A7 phase.

Our experimental findings can be benchmarked against first principles density functional theory (DFT) calculations of the structural, electronic, and optical properties under pressure. A first-principles description of the BP electronic properties across the semiconductor-metal transition is challenging for local DFT exchange-correlation functionals, which predict a metallic ground state at ambient conditions. In order to reproduce the small band gap at ambient conditions, we used the Tran-Blaha\cite{TB} meta-GGA exchange-correlation potential in DFT calculations~\cite{VASP}, which is quite reliable in describing small gap $sp$ systems. The calculated zero-pressure band gap is 2000 cm$^{-1}$ for the experimental, ambient-pressure lattice parameters, in agreement with the ambient pressure optical conductivity (see Fig. 3b), and previous experimental reports \cite{ehlen16,kim15}. More details about the DFT calculations are reported in the supplementary information. 
By using experimental structural information as a function of pressure from Ref. \onlinecite{Kikeg83}, we calculated the plasma frequencies reported and compared with the experiment in Fig.\ref{Fig3}e.  Although the qualitative behavior is well reproduced, the results consistently overestimate the experimental values by a factor of 3. 
The onset of the Lifshitz transition is theoretically predicted at 2.1 GPa, close to, but slightly higher than the 1.5 GPa experimental value. This small discrepancy is likely due to defects and to the anomalous temperature dependence of the band structure\cite{villegas16,ehlen16}.

Above 3 GPa the theoretical plasma frequency calculated within the A17 structural phase increases linearly with pressure up to 8.5 GPa as experimentally observed (Fig. \ref{Fig3}e). Considering the structural phase transition in the A7 phase, we found that the theoretical plasma frequencies (green triangles in Fig. \ref{Fig3}e) are significantly enhanced with respect to the A17 case. Interestingly, the experimental values can match this leap if one considers the $\omega_{p,{\textrm{Drude+MIR}}}$ plasma frequency (green circles in Fig. \ref{Fig3}e) instead of $\omega_p$. The qualitative agreement between experimental data and theoretical calculations within this large pressure range and in different structural phases indicates that the midinfrared band can be attributed to partially localized charge carriers emerging in the A7 phase and coalescing in a Drude term at higher pressures. This spectral feature, likely related to strong interactions, is intriguing and deserves further study.

In conclusion, we presented the first direct optical identification of a pressure-induced Lifshitz transition in elemental BP ($P_L=1.5$ GPa). The key spectral feature associated with this transition is a discontinuity in the pressure-dependent carrier density that can be attributed to the emergence of a plasma of massless Dirac carriers. The character of the Lifshitz transition has been confirmed through comparison with DFT calculations that provided an excellent description of the experimental plasma frequencies. The Dirac plasma frequency increases linearly with pressure, well into the A17-A7 mixed phase up to about 8 GPa. The onset of the A7 structural phase triggers the delocalization of a significant portion of charge carriers which become indistinguishable from the Dirac carriers when entering into the sc phase above 9 GPa. Our work in a clean, controlled elemental system will serve as a useful guide to identify optical signatures of a Lifshitz transition in more complicated systems, like hole-doped cuprates and iron pnictides, and will lead to a deeper understanding of this fascinating physical phenomenon.

\section*{Acknowledgements}
The authors wish to thank A. Cavalleri for providing the DAC for lower pressure measurements, and L. Ortenzi and E. Cappelluti for preliminary discussions on the experimental data. B.J. acknowledges the IISc-ICTP fellowship from IISc Bangalore and ICTP Trieste.

\end{document}